\documentclass[prl,showpacs,twocolumn,amssymb]{revtex4}
\begin{document}

\title{Magnetic Fields Boosted by Gluon Vortices in Color Superconductivity}
\author{Efrain J. Ferrer}
\author{Vivian de la Incera}
\affiliation{Department of Physics, Western Illinois University,
Macomb, IL 61455, USA}

\begin{abstract}

We investigate the effects of an external magnetic field in the
gluon dynamics of a color superconductor with three massless quark
flavors. In the framework of gluon mean-field theory at asymptotic
densities, we show that the long-range component $\widetilde{H}$
of the external magnetic field that penetrates the CFL phase
produces an instability when its strength becomes larger than the
Meissner mass of the charged gluons. As a consequence, the
magnetic field causes the formation of a vortex state
characterized by the condensation of charged gluons and the
creation of magnetic flux tubes. Inside the flux tubes the
magnetic field is stronger than the applied one. This
antiscreening effect is connected to the anomalous magnetic moment
of the gluon field. We suggest how this same mechanism could serve
to remove the chromomagnetic instabilities existing in gapless
color superconductivity.

\end{abstract}
\maketitle It is quite plausible \cite{reviews} that color
superconductivity (CS) occurs in the inner regions of neutron
stars, hence affecting the structure and properties of these
compact objects. At the same time, it is known that strong
magnetic fields exist in the surface of compact stars, reaching
values of orders $10^{14} - 10^{16}$ G in the case of magnetars
\cite{magnetars}. Comparing magnetic and gravitational energies it
has been shown \cite{Lai:2000at} that the physical upper limit for
the total neutron star magnetic field is $B \sim 10^{18}$ G. If
quark stars indeed exist and are self-bound rather than
gravitational-bound this upper limit can be even higher.
Therefore, the investigation of CS in the presence of a magnetic
field is of main interest for possible astrophysical applications.

It is well established by now that at densities high enough to
neglect the strange quark mass, the ground state of three-flavor
quark matter corresponds to the color-flavor locked (CFL) phase
\cite{alf-raj-wil-99/537}. In this phase, quarks form spin-zero
Cooper pairs in the color-antitriplet, flavor-antitriplet
representation. An important feature of spin-zero color
superconductivity is that although the color condensate has
non-zero electric charge, the linear combination
$\widetilde{A}_{\mu}=\cos{\theta}\,A_{\mu}+\sin{\theta}\,G^{8}_{\mu}$
of the photon $A_{\mu}$ and the gluon $G^{8}_{\mu}$ remains
massless \cite{alf-raj-wil-99/537, alf-berg-raj-NPB-02}, so it
behaves as an in-medium electromagnetic field, while the
orthogonal combination
$\widetilde{G}_{\mu}^8=-\sin{\theta}A_{\mu}+\cos{\theta}\,G^{8}_{\mu}$
is massive. In the CFL phase the mixing angle $\theta$ is small
thus the penetrating "electromagnetic" field is mostly formed by
the original photon with only a small gluon admixture.

Even though gluons are electrically neutral, in the CFL phase they
can interact with the modified (also called rotated) photon
through their $\widetilde{Q}$ charges:
\begin{equation} \label{table}
\begin{tabular}{|c|c|c|c|c|c|c|c|c|}
  \hline
  $G_{\mu}^{1}$ & $G_{\mu}^{2}$ & $G_{\mu}^{3}$ & $G_{\mu}^{+}$ & $G_{\mu}^{-}$ & $I_{\mu}^{+}$ & $I_{\mu}^{-}$ & $\widetilde{G}_{\mu}^{8}$ \\
  \hline
  0 & 0 & 0 & 1 & -1 & 1 & -1 & 0 \\
  \hline
\end{tabular} \ ,
\end{equation}
given in units of $\widetilde{e} = e \cos{\theta}$. The
$\widetilde{Q}$-charged fields in (\ref{table}) correspond to the
combinations $G_{\mu}^{\pm}\equiv\frac{1}{\sqrt{2}}[G_{\mu}^{4}\pm
iG_{\mu}^{5}]$ and
$I_{\mu}^{\pm}\equiv\frac{1}{\sqrt{2}}[G_{\mu}^{6}\pm
iG_{\mu}^{7}]$. If the color superconductor is penetrated by a
sufficiently strong rotated magnetic field, the structure and
magnitude of the CFL gap are modified giving rise to a new phase
called Magnetic CFL (MCFL) \cite{MCFL}. The MCFL \cite{MCFL} was
obtained in the context of an effective NJL model in which the
interactions between the $\widetilde{Q}$-charged gluons and the
magnetic field were ignored. In this letter, we are interested in
extending those previous investigations by considering the impact of
an external magnetic field on the gluon dynamics. We will show that
at high enough magnetic field there is an inhomogeneous condensate
of the $\widetilde{Q}$-charged gluons that antiscreens the magnetic
field due to the anomalous magnetic moment of these spin-1
particles. As a consequence, this condensate does not give a mass to
the $\widetilde{Q}$ photon, but instead it amplifies the applied
$\widetilde{Q}$ magnetic field (for the analogous behavior of W
bosons in a high magnetic field, see \cite{Olesen}).

We will work in the weak coupling region of QCD, assuming that the
external magnetic field is much smaller than the square of the
chemical potential, so that the effects of the field on the gap
structure and magnitude can be neglected. In this case, it is safe
to assume that the superconductor is in the CFL phase.

We start from the gauge field sector of the mean-field effective
action in the CFL phase at asymptotic densities. This effective
action, which includes the one-loop quark contribution obtained
after integrating out the Nambu-Gorkov fields, can be written as
\begin{eqnarray}
\label{Eff-Act} \Gamma_{eff} & = & - \frac{1}{4}\int d^{4}x
[(G_{\mu \nu}^{a})^{2}+(f_{\mu \nu})^{2}]+\int d^{4}x
\textsl{L}_{g}(x) \nonumber
 \\
& - & \frac{1}{2} \int d^{4}x d^{4}y G_{\mu}^{a}(x) \Pi_{\mu
\nu}^{a b}(x,y) G_{\nu}^{b}(y)\nonumber
\\
& - & \frac{1}{2} \int d^{4}x d^{4}y A_{\mu}(x) \Pi_{\mu \nu}
(x,y) A_{\nu}(y)
\end{eqnarray}
where $\textsl{L}_{g}$ is the gauge fixing Lagrangian density,
$G_{\mu \nu}^{a}$ and $f_{\mu \nu}$ are the gluon and
electromagnetic field strength tensors respectively, and $\Pi_{\mu
\nu}^{a b}$ and $\Pi_{\mu \nu}$ are their corresponding
polarization operators in the CFL phase \cite{Pi-CFL}. No mixing
between the gauge fields and the Nambu-Goldstone mesons is
included because this mixing can be eliminated using the 't Hooft
gauge \cite{Rischke}.

Since we are interested in the CFL-system response to a constant
external magnetic field, we can concentrate our attention into the
$\widetilde{Q}$-charged gluon part of (\ref{Eff-Act}) in the
presence of an external rotated magnetic field $\widetilde{H}$.
Without loss of generality, the analysis can be done for one of
the two sets of charged fields, say $G_{\mu}^{\pm}$. For the
static response, one only needs the leading contribution of the
polarization operators \cite{Pi-CFL} of the
$\widetilde{Q}$-charged gluons in the infrared limit ($p_{0}=0,
|\overrightarrow{p}|\rightarrow 0$): $\label{Pi} \Pi_{\mu \nu}^{a
b}(x,y)= [m_{D}^{2}\delta_{\mu0}\delta_{\nu0} +
m_{M}^{2}\delta_{\mu i}\delta_{\nu i}]\delta (x-y)\delta^{ab}$,
with a,b=4,...,7. The Debye ($m_{D}$) and Meissner ($m_{M}$)
masses are given by $m_{D}^{2} =\sigma m_{g}^{2}$, $m_{M}^{2}
=\sigma m_{g}^{2}/3$, with $\sigma =(21-8\ln 2)/18$, and
$m_{g}^{2}=g^2(\mu^{2}/2\pi^{2})$. We neglect the possible
corrections produced by the applied field in the infrared masses
since for fields weak compared to the quark chemical potential, it
is a second order effect. The effective action for the fields
$G_{\mu}^{\pm}$ in the background of the $\widetilde{H}$ external
field becomes
\begin{eqnarray}
\label{Eff-Act-2} &\Gamma_{eff}^{c}=  \int d^{4}x
\{-\frac{1}{4}(\widetilde{f}_{\mu
\nu})^{2}-\frac{1}{2}|\widetilde{\Pi}_{\mu}G_{\nu}^{-}-\widetilde{\Pi}_{\nu}G_{\mu}^{-}|^{2}&
\nonumber
 \\
& -  [(m_{D}^{2} \delta_{\mu 0} \delta_{\nu 0}+ m_{M}^{2}
\delta_{\mu i} \delta_{\nu i})+ i\widetilde{e}\widetilde{f}_{\mu
\nu}] G_{\mu}^{+}G_{\nu}^{-}& \nonumber
 \\
 & +
\frac{g^2}{2}[(G^{+}_{\mu})^{2}(G^{-}_{\nu})^{2}-(G^{+}_{\mu}G^{-}_{\mu})^{2}]+\frac{1}{\lambda}G_{\mu}^{+}\widetilde{\Pi}_{\mu}\widetilde{\Pi}_{\nu}G_{\nu}^{-}
\},&
\end{eqnarray}
where the last term in (\ref{Eff-Act-2}) comes from the
$\textsl{L}_{g}$ term in (\ref{Eff-Act}), taken in the 't Hooft
gauge with arbitrary gauge fixing parameter $\lambda$,
$\widetilde{\Pi}_{\mu}=\partial_{\mu}
-i\widetilde{e}\widetilde{A}_{\mu}$, and $\widetilde{f}_{\mu
\nu}=\widetilde{f}_{1 2}^{ext}=\widetilde{H}$, that is, we take
the external rotated magnetic field along the third spatial
direction. The effective action (\ref{Eff-Act-2}) is nothing but
the characteristic effective action of a spin-1 charged boson in a
magnetic field \cite{emilio}. As known, due to the anomalous
magnetic moment term ($i\widetilde{e}\widetilde{f}_{\mu
\nu}G_{\mu}^{+}G_{\nu}^{-} $), when the field surpasses a critical
value, one of the modes of the charged gauge field becomes
tachyonic (this is the well known "zero-mode problem" found in the
presence of a magnetic field for Yang-Mills fields
\cite{zero-mode}, for the $W^{\pm}_{\mu}$ bosons in the
electroweak theory \cite{Skalozub, Olesen}, and even for
higher-spin fields in the context of string theory
\cite{porrati}). The tachyonic mode can be easily found from
(\ref{Eff-Act-2}) diagonalizing the mass matrix of the field
components ($G^{\pm}_{1}, G^{\pm}_{2}$)
\begin{equation}
\left(
\begin{array}{cc}
m_{M}^{2}& i\widetilde{e}\widetilde{H} \\
- i\widetilde{e}\widetilde{H}& m_{M}^{2}
 \label{mass-matrx}
\end{array} \right) \rightarrow
\left(
\begin{array}{cc}
m_{M}^{2}+\widetilde{e}\widetilde{H}& 0 \\
0& m_{M}^{2}-\widetilde{e}\widetilde{H}
 \label{mass-matrx}
\end{array} \right)
\end{equation}
Clearly, above the critical field $\widetilde{e}\widetilde{H}_{C}=
m_{M}^2$ the lowest mass mode in (\ref{mass-matrx}) becomes
tachyonic, with corresponding eigenvector of amplitude $G$ in the
$(1,i)$ direction for $G^{-}$ ($G^{\ast}$ in the $(1,-i)$
direction for $G^{+}$). We emphasize that the $\widetilde{H}$
field producing the instability is the in-medium long-range field
that does not suffer the Meissner effect in the CFL phase.

Similarly to other spin-1 theories with magnetic instabilities
\cite{Olesen}-\cite{zero-mode}, the solution of the zero-mode
problem leads to the restructuring of the ground state through the
formation of a gauge field condensate $G$, as well as an induced
magnetic field
$\widetilde{\textbf{B}}=\nabla\times\widetilde{\textbf{A}}$ due to
the backreaction of the G condensate on the rotated electromagnetic
field.

The condensate solutions can be found by minimizing the Gibbs free
energy density
$\mathcal{G}_{c}=\mathcal{F}-\widetilde{H}\widetilde{B}$,
($\mathcal{F}$ is the free energy density), with respect to $G$
and $\widetilde{B}$. Since an applied field
$\widetilde{H}>\widetilde{H}_{C}$ in the third direction develops
an instability in the $(x,y)$-plane for the eigenmode $G(1,i)$, it
is reasonable to make the ansatz $G_{1}^{-}=-iG_{2}^{-}=G(x,y)$,
$G_{3}^{-}=G_{0}^{-}=0$ and $G_{i}^{+}=(G^{-}_{i})^{\ast}$.
Considering this ansatz and fixing the gauge parameter to
$\lambda=1$ in the action (\ref{Eff-Act-2}) we obtain for the
Gibbs free energy density in the $G$-condensate phase
\begin{eqnarray}
\label{Gibbs} \mathcal{G}_{c} =\mathcal{F}_{n0}
-2G^{\dag}\widetilde{\Pi}^{2}
G-2(2\widetilde{e}\widetilde{B}-m_{M}^{2})|G|^{2}+2g^{2}|G|^{4}\nonumber
 \\
+ \frac{1}{2}\widetilde{B}^{2}-\widetilde{H}\widetilde{B}\qquad
\qquad \qquad \qquad \qquad \qquad \qquad
\end{eqnarray}
where $\mathcal{F}_{n0}$ is the system free energy density in the
normal-CFL phase ($G=0$) at zero applied field.

Using (\ref{Gibbs}) the minimum equations for the condensate $G$
and induced field $\widetilde{B}$ respectively are
\begin{equation}
\label{G-Eq} -\widetilde{\Pi}^{2}
G-(2\widetilde{e}\widetilde{B}-m_{M}^{2})G+2g^{2}|G|^{2}G=0,
\end{equation}
\begin{equation}
\label{B-Eq} 2\widetilde{e} |G|^{2}-\widetilde{B}+\widetilde{H}=0
\end{equation}
Identifying $G$ with the complex order parameter, Eqs.
(\ref{G-Eq})-(\ref{B-Eq}) become analogous to the Ginzburg-Landau
equations for a conventional superconductor except for the
$\widetilde{B}$ contribution in the second term in (\ref{G-Eq})
and the sign of the first term in (\ref{B-Eq}). The origin of both
terms can be traced back to the anomalous magnetic moment term in
the action of the charged gluons. Notice that because of the
different sign in the first term of (\ref{B-Eq}), contrary to what
occurs in conventional superconductivity, the resultant field
$\widetilde{B}$ is stronger than the applied field
$\widetilde{H}$. Thus, when a gluon condensate develops, the
magnetic field will be antiscreened and the CS will behave as a
paramagnet. The antiscreening of a magnetic field has been also
found in the context of the electroweak theory for magnetic fields
$eH \geq M_{W}^{2}$ \cite{Olesen}. Just as in the electroweak
case, the antiscreening of the CS is a direct consequence of the
asymptotic freedom of the underlying theory \cite{Olesen, Hughes}.

The main goal of this paper will be to investigate the qualitative
features of the new phase with the gluon condensate at
$\widetilde{H}\simeq \widetilde{H}_{C}$. A starting point in this
direction will be to investigate the sign of the condensation energy
\begin{eqnarray}
\label{Surf-Tens-1} \alpha_{nc}=\int_{-\infty}^{\infty}
[\mathcal{G}_{c}-\mathcal{G}_{n}]dx =
\int_{-\infty}^{\infty}\{-2G^{\dag}\widetilde{\Pi}^{2}
G+\frac{1}{2}(\widetilde{B}-\widetilde{H}_{C})^{2} \nonumber
\\
-2(2\widetilde{e}\widetilde{B}-m_{M}^{2})|G|^{2}+2g^{2}|G|^{4}\}dx
\qquad
\end{eqnarray}
with $\mathcal{G}_{n}=\mathcal{F}_{n0}-\widetilde{H}_{C}^{2}/2$
being the Gibbs free energy density of the normal-CFL phase in the
external rotated magnetic field $\widetilde{H}_{C}$. Considering
that just below $\widetilde{H}_{C}$ the system is in the normal-CFL
phase, we should look for the possibility of nucleation at
$\widetilde{H}_{C}$ of an inhomogeneous condensate satisfying
asymptotic boundary conditions that correspond to the normal-CFL
phase, i.e. $G(x\rightarrow\pm\infty) = 0$. Using the
Ginzburg-Landau equations (\ref{G-Eq})-(\ref{B-Eq}) we can rewrite
(\ref{Surf-Tens-1}) as
\begin{equation}
\label{Surf-Tens-2}
\alpha_{nc}=\int_{-\infty}^{\infty}(1-\frac{g^{2}}{\widetilde{e}^{2}})
\frac{(\widetilde{B}-\widetilde{H}_{C})^{2}}{2} dx
\end{equation}
Because of the hierarchy between the strong ($g$) and
electromagnetic ($\widetilde{e}$) interactions, we have that
$g^2/\widetilde{e}^2
> 1$ and hence $\alpha_{nc}<0$, indicating that the
nucleation of an inhomogeneous condensate $G$ is energetically
favored.

To obtain the explicit expression for the condensate solution, we
can follow Abrikosov's approach \cite{Abrikosov} to type II metal
superconductivity for the limit situation when the applied field is
near the critical value $H_{c2}$. Close to $\widetilde{H}_{C}$ the
amplitude of the condensate $G$ should be very small and we can
neglect the nonlinear term in Eq. (\ref{G-Eq}). Similarly, the
$G^{2}$ term in (\ref{B-Eq}) can be neglected yielding
$\widetilde{B}\approx \widetilde{H}_{C}$. As a consequence, Eqs.
(\ref{G-Eq})-(\ref{B-Eq}) decouple and $G$ can be found from
\begin{equation}
\label{Vortex-Eq} [\partial_{j}^{2}-\frac{4\pi
i}{\widetilde{\Phi}_{0}}\widetilde{H}_{C}x\partial_{y}-4\pi^{2}\frac{\widetilde{H}_{C}^{2}}{\widetilde{\Phi}_{0}^{2}}x^{2}+\frac{1}{\xi^{2}}]G=0,
\quad j=x,y
\end{equation}
where we used the gauge $\widetilde{A}_{2}=\widetilde{H}_{C}x_{1}$
and defined $\widetilde{\Phi}_{0}\equiv2\pi/\widetilde{e}$, and
$\xi^{2}\equiv1/(2\widetilde{e}\widetilde{H}_{C}-m_{M}^{2})=1/m_{M}^{2}$.

A solution of Eq. (\ref{Vortex-Eq}) can be proposed in the form
\begin{equation}
\label{Sol-1} G(x,y)=e^{-ik_{y}y}f(x)
\end{equation}
with $f(x)$ satisfying the differential equation
\begin{equation}
\label{Sol-2} -f''(x)+4\pi^2
\frac{\widetilde{H}_{C}^{2}}{\widetilde{\Phi}_{0}^{2}}(x-x_{k})^{2}f=\frac{1}{\xi^{2}}f,
\end{equation}
and vanishing at $x\rightarrow\pm\infty$. The parameter $x_{k}$ is
$x_{k}\equiv k_{y}\widetilde{\Phi}_{0}/2\pi \widetilde{H}_{C}$. Eqs.
(\ref{Vortex-Eq}) and (\ref{Sol-2}) are formally identical to those
found for the Abrikosov's vortex state near the critical field
$H_{c2}$. The parameter $\xi$ in (\ref{Vortex-Eq}) is playing the
role of the coherence length, i.e. the characteristic length for the
variation of $G$. Increasing the magnetic field strength makes $\xi$
to decrease. The solution of (\ref{Vortex-Eq}), (\ref{Sol-2}) is
\begin{equation}
\label{Sol-3} G_{k}=\exp {[-iky]}
\exp{[-\frac{(x-x_{k})^2}{2\xi^{2}}]}
\end{equation}
where $k\equiv k_{y}$. From the experience with conventional type II
superconductivity  \cite{Tinkham} it is known that the inhomogeneous
condensate solutions prefer periodic lattice domains to minimize the
energy. Then, putting on periodicity in the $y$-direction with
period $\Delta y= b$ restricts the values of $k$ to a discrete set
$k=2\pi n/b$,   $n=1,2,..$. This condition implies that we have an
infinite set of discrete solutions, and because Eqs.
(\ref{Vortex-Eq}) and (\ref{Sol-2}) are linear differential
equations, the general solution should be given by their
superposition $G(x,y)=\sum C_{n}G_{n}$. The superposition of all
these Gaussian solutions centered at different $x_{n}$ constitutes
the vortex state needed to solve the instability in the whole space.
On the other hand, the discrete values of $k$ imply periodicity in
$x$, since the Gaussian solutions $G_{n}$ are located at
$x_{n}=\frac{k_{n}\widetilde{\Phi}_{0}}{2\pi
\widetilde{H}_{C}}$=$\frac{ n\widetilde{\Phi}_{0}}{b
\widetilde{H}_{C}}$. Hence, assuming that all $G_{n}$ enter with
equal weight, the periodicity length in the $x$-direction is $\Delta
x=\frac{\widetilde{\Phi}_{0}}{b \widetilde{H}_{C}}$. Therefore, the
magnetic flux through each periodicity cell in the vortex lattice is
quantized $\label{Flux} \widetilde{H}_{C}\Delta x \Delta
y=\widetilde{\Phi}_{0}$, with $\widetilde{\Phi}_{0}$ being the flux
quantum per unit vortex cell. In this semi-qualitative analysis we
considered Abrikosov's ansatz of a rectangular lattice (i.e. all the
coefficients $C_{n}$ being equal), but the lattice configuration
should be carefully determined from a minimal energy analysis. For
the rectangular lattice, we see that the area of the unit cell is
$A=\Delta x \Delta y=\widetilde{\Phi}_{0} /\widetilde{H}_{C}$, so
decreasing with $\widetilde{H}$.

Substituting with Eq.(\ref{Sol-3}) back into (\ref{B-Eq}) to find
a correction to the linear solution for the induced field
$\widetilde{B}$, one easily sees that the induced field, while
being homogeneous in the $z$-direction, becomes inhomogeneous in
the $(x,y)$-plane since it depends on the condensate $G$, which
has periodicity on that plane. Therefore $\widetilde{B}$ forms a
fluxoid along the $z$-direction creating a nontrivial topology on
the $(x,y)$ plane. Notice that the magnetic flux found in the
present paper is qualitatively different from the one found in
Ref. \cite{Balachandran}, where the flux was related to the
massive gauge field $\widetilde{G}_{\mu}^{8}$. In
\cite{Balachandran} the flux was due to the existence of
nontrivial loops in the order parameter (diquark condensate) space
that originated in the transition between the unpaired phase and
the CFL phase.

In conclusion, at low $\widetilde{H}$ field, the CFL phase is an
insulator, and the $\widetilde{H}$ field just penetrates through it.
At sufficiently high $\widetilde{H}$, the condensation of $G^{\pm}$
is triggered inducing the formation of a lattice of magnetic flux
tubes and breaking the translational and remaining rotational
symmetries. Contrary to the situation in conventional type-II
superconductors, where the applied field only penetrates through the
flux tubes and with a smaller strength, the vortex state found in
this paper has the peculiarity that outside the flux tube the
applied field $\widetilde{H}$ totally penetrates the sample, while
inside the tubes the magnetic field becomes larger than
$\widetilde{H}$. This antiscreening behavior is similar to that of
the electroweak system at high magnetic field \cite{Olesen}. Notice
that as the $\widetilde{Q}$ photons remain massless in the presence
of the condensate $G$, the $\widetilde{U}(1)_{em}$ symmetry remains
unbroken.

A rough estimate of the critical field that produces the magnetic
instability at the scale of baryon densities typical of
neutron-star cores ($\mu \simeq 200-400 MeV$, $\alpha_{s}(\mu)
\simeq 1/3$) gives $\widetilde{H}_{C}\simeq 9.5\times
10^{16}G-3.8\times 10^{17}G$. Although these are significantly
high magnetic fields, they cannot be ruled out as acceptable
values for the neutron star core.

We would like to call attention to a possible connection between
our results and the problem of the instabilities \cite{Igor,
Fukushima} in gapless CS \cite{Alford}. As known, a color
superconductor can develop chromomagnetic instabilities even in
the absence of an external magnetic field. These instabilities may
appear after imposing electrical and color neutralities and
$\beta$ equilibrium conditions, and at densities where the $s$
quark mass $M_{s}$ becomes a relevant parameter. As found first in
$g2SC$ \cite{Igor}, and then also in $gCFL$ \cite{Fukushima}, some
charged gluons typically become tachyonic at the onset of the
gapless phase. From the outcome of this paper it results clear
that although the imaginary Meissner mass of these charged gluons
in the gapless phase is not triggered by any external magnetic
field, the mechanism we investigated can work here to remove the
instability of the gapless phase. In this case we could use the
same effective action (\ref{Eff-Act-2}) as a toy model to describe
the instabilities of the charged gluons in gapless CS. The main
difference will be to consider a negative square Meissner mass
from the beginning and to assume that although no external
magnetic field is present ($\widetilde{H}=0$), the induced
magnetic field $\widetilde{B}$ can be different from zero (see
Eq.(8)) since it is nothing but the rotated electromagnetic field
generated by the gluon condensate. Then the instability will be
removed through the spontaneous breaking of the spatial rotational
symmetry $SO(3)\rightarrow SO(2)$ due to vortex nucleation and the
induction of a rotated magnetic field perpendicular to the vortex
planes. Notice that the new equation for the magnetic field,
$2\widetilde{e} G^{2}-\widetilde{B}=0$, can have a nontrivial
solution thanks to the antiscreening effect produced by the
anomalous magnetic moment contribution $2\widetilde{e} G^{2}$.
This means that the induction of a $\widetilde{B}$ field, together
with the topological modification of the medium that goes with it,
can be directly linked to the asymptotic freedom of the theory. It
remains to be seen what would be the relative field orientation
between neighboring domains that minimizes the system free-energy,
and whether the generation of $G$ and $\widetilde{B}$ alone is
enough to remove the instability of the neutral gluon modes in the
gapless phase. If our arguments are corroborated, a color
superconducting core could provide a new mechanism to generate and
amplify the magnetic fields of compact stars. The idea of using
gluon condensates to solve the chromomagnetic instabilities in CS
has been also explored in Refs. \cite{miransky}.

It is worth to mention that if the ground state is
self-consistently found, by minimizing the free-energy with
respect to all the possible condensates: gluon and diquark
condensates and the induced rotated magnetic field, the
inhomogeneous gluon condensate would produce a backreaction into
the diquark condensate making it inhomogeneous too and quite
likely with the same periodicity of the gluon vortex solution.

{\bf Acknowledgments:} We are grateful to K.Fukushima, C.Manuel,
 L.McLerran, R.D.Pisarski and  I.A.Shovkovy for valuable comments.


\begin{thebibliography}{}

\bibitem{reviews}
  K.~Rajagopal and F.~Wilczek,
  %``The condensed matter physics of QCD,''
 hep-ph/0011333; M.~Alford, Ann.\ Rev.\ Nucl.\ Part.\ Sci.\ {\bf 51}, 131 (2001);
G.~Nardulli,  Riv.\ Nuovo Cim.\  {\bf 25N3}, 1 (2002); T.
Sch\"afer, hep-ph/0304281; D.~H.~Rischke, Prog.\ Part.\ Nucl.\
Phys.\ {\bf 52}, 197 (2004); I.~A.~Shovkovy, nucl-th/0410091.


\bibitem{magnetars}
  C.~Thompson and R.~C.~Duncan,
  %``The Soft gamma repeaters as very strongly magnetized neutron stars. 2.
  %Quiescent neutrino, x-ray, and Alfven wave emission,''
  ApJ {\bf 473}, 322 (1996).
  %%CITATION = ASJOA,473,322;%%

%\cite{Lai:2000at}
\bibitem{Lai:2000at}
I. Fushiki, E. H. Gudmundsson, and C.J. Pethick, ApJ 342, 958
(1989); T.A. Mihara, et. al., Nature (London) 346, 250 (1990);
D.~Lai, Rev.\ Mod.\ Phys. {\bf 73}, 629 (2001).
  %``Matter in Strong Magnetic Fields,''
%  arXiv:astro-ph/0009333.
  %%CITATION = ASTRO-PH 0009333;%%

\bibitem{alf-raj-wil-99/537}
M. Alford, K. Rajagopal and F. Wilczek, Nucl. Phys. B
\textbf{537}, 443 (1999).

\bibitem{alf-berg-raj-NPB-02}
M. Alford, J. Berges, and K. Rajagopal, Nucl. Phys. B
\textbf{571}, 269 (2000); E. V. Gorbar, Phys. Rev. D \textbf{62},
014007 (2000).

\bibitem{MCFL}
   E.~J.~Ferrer, V.~de la Incera and C.~Manuel,
  %``Magnetic color flavor locking phase in high density QCD,''
  Phys.\ Rev.\ Lett.\  {\bf 95}, 152002 (2005); Nucl. Phys. B \textbf{747}, 88 (2006).
  %%CITATION = HEP-PH 0503162;%%

\bibitem{Olesen}J. Ambjorn and P. Olesen, Nucl. Phys. B \textbf{315},
606 (1989); Phys. Lett. B \textbf{218}, 67 (1989).

\bibitem{Pi-CFL}
D. T. Son and M. A. Stephanov, Phys. Rev. D \textbf{61}, 074012
(2000); M. Rho, E. Shuryak, A. Wirzba, and I. Zahed, Nucl. Phys. A
\textbf{676}, 273 (2000); S. R. Beane, P. F. Bedaque, and M. J.
Savage, Phys. Lett. \textbf{483}, 131 (2000); K. Zarembo, Phys.
Rev. D \textbf{62}, 054003 (2000); D. H. Rischke, Phys. Rev. D
\textbf{62}, 054017 (2000); D.F. Litim and C. Manuel, Phys. Rev. D
\textbf{64}, 094013 (2001); A. Schmitt, Q. Wang, and D. H.
Rischke, Phys. Rev. D \textbf{69}, 094017 (2004).

\bibitem{Rischke}D. H. Rischke, and I. A. Shovkovy, Phys. Rev. D \textbf{66}, 054019 (2002).

\bibitem{emilio}
E. Elizalde and E. J. Ferrer, and V. de la Incera, Ann. of Phys.,
{\bf 295}, 33 (2002); \prd {\bf 70}, 043012 (2004).
%%CITATION = HEP-PH 0404234;%%
%%CITATION = HEP-PH 0007033;%%

\bibitem{zero-mode}V. V. Skalozub, Sov. J. Nucl. Phys. \textbf{23},
113 (1978); N. K. Nielsen and P. Olesen, Nucl. Phys. B
\textbf{144}, 376 (1978).

\bibitem{Skalozub}V. V. Skalozub, Sov. J. Nucl. Phys. \textbf{43}, 665 (1986);
\textbf{45}, 1058 (1987).


\bibitem{porrati}S. Ferrara and M. Porrati, Mod. Phys. Lett. A \textbf{8},
2497 (1993); E. J. Ferrer and V. de la Incera, Int. Jour. of Mod.
Phys. A \textbf{11}, 3875 (1996).

\bibitem{Hughes}R. Hughes, Phys. Lett. B \textbf{97}, 246 (1980); J. Ambjorn and P. Olesen, Int. Journ. Mod. Phys. A \textbf{5}, 4525 (1990).

\bibitem{Abrikosov}A. A. Abrikosov, Sov. Phys. JEPT \textbf{5}, 1174
(1957).

\bibitem{Tinkham} M. Tinkham, \textit{Introduction
to Superconductivity}, Robert E. Krieger Publ. Com., 1985. (See
Section 4-11).

\bibitem{Balachandran} A.P. Balachandran, S. Digal, and T.
Matsuura, \prd {\bf 73}, 074009 (2006).

\bibitem{Igor} M. Huang and I. A. Shovkovy, \prd {\bf 70}, 051501 (2004); {\bf 70}, 094030 (2004).

\bibitem{Fukushima}R.~Casalbuoni, R.~Gatto, M. Mannarelli, G.~Nardulli and M.~Ruggieri, Phys. Lett. B \textbf{605}, 362 (2005);  \textbf{615}, 297(E) (2005), M. Alford and Q. H. Wang, J. Phys. G \textbf{31}, 719 (2005); K. Fukushima, Phys. Rev. D \textbf{72}, 074002 (2005).


\bibitem{Alford}I. A. Shovkovy, and M. Huang, Phys. Lett. B \textbf{564}, 205 (2003); M. Alford, C. Kouvaris, and K. Rajagopal, Phys. Rev. Lett. \textbf{92}, 222001
(2004).

\bibitem{miransky} E.~V.~Gorbar, M~Hashimoto, and V.~A.~Miransky, Phys. Lett. B \textbf{632}, 305 (2006); E.~V.~Gorbar, Junji Jia, and V.~A.~Miransky, Phys. Rev. D \textbf{73},
045001 (2006); K. Fukushima, hep-ph/0603216.

\end{thebibliography}
\end{document}